\documentstyle[12pt]{article}
\begin{document}
\begin{center}
\Large\bf Hybrid Quarkonium Masses up to the order of ${\cal O}(1/m_Q)$
\end{center}
\vspace{0.5cm}
\begin{center}
{ Shi-Lin Zhu}\\\vspace{3mm}
{Institute of Theoretical Physics \\
Academia Sinica, P.O.Box 2735\\
Beijing 100080, China\\
FAX: 086-10-62562587\\
TEL: 086-10-62569358\\
E-MAIL: zhusl@itp.ac.cn}
\end{center}
\vspace{1.0cm}
\begin{abstract}
We have calculated the binding energy of the hybrid quarkonium
up to the order of ${\cal O}(1/m_Q)$ and found their decay constants
scale like $m_Q^{5\over 4}$ as $m_Q\to \infty$. 
The $0^{--}$ and $0^{++}$ hybrid quarkonium is exactly degenerate 
in the limit $m_Q\to \infty$ while the ${\cal O}(1/m_Q)$ correction renders
the $0^{--}$ mass lower than that for $0^{++}$. The $1^{-+}$ and $1^{+-}$
hybrid is nearly degenerate and lies $0.7$ GeV lower than the $0^{--}$. 
If we use $m_b=4.8$GeV the mass of the $1^{-+}$ $b\bar b g$ state is 
$(10.75\pm 0.20)$ GeV, which is the lightest exotic hybrid quarkonium.
\end{abstract}



\pagenumbering{arabic}
\vskip 1.0cm

Quantum Chromodynamics
(QCD) is believed to be the correct theory of strong interaction. In the 
massless light quark limit, QCD has exact $SU_L(3)\times SU_R(3)$ symmetry. 
On the other hand, QCD has spin-flavor symmetry in the heavy quark limit
$m_Q\to \infty$. In both cases we have an effective theory: the chiral 
perturbation theory ($\chi$PT) and the heavy quark effective theory (HQET).
$\chi$PT is very useful for the low energy dynamics of strong interaction,
while HQET provides a consistent framework to study hadrons containing one 
heavy quark. Symmetry plays a fundamental role in particle physics. It is
important to explore novel features of the exotic system $(Q\bar Q g )$ with
two heavy quarks in the limit $m_Q\to\infty$.

From quark model we know that 
a $q{\bar q}$ meson with orbital angular momentum l and total spin s must have 
$P=(-1)^{l+1}$ and $C=(-1)^{l+s}$. Thus a resonance with $J^{PC}=0^{--}, 
0^{+-}, 1^{-+}, 2^{+-}, \cdots$ must be exotic. Such a state could be a 
multiquark state or gluonic excitation such as hybrids, glueballs.
Very recently there appears experimental evidence for a light $J^{PC}=1^{-+}$ exotic
\cite{e852,cb,gams,kek,ves}. The emergence of evidence for hybrids 
indicates the presence of dynamical glue in QCD.

The traditional theoretical approach to the gluonic excitations in mesons
falls into two classes: the string or flux tube model and the constituent
glue model with the glue confined by a bag or potential.
In contrast, QCD sum rule (QSR) starts from the first principle 
like the lattice gauge theory and incorporates the nonperturbative 
effects via various condensates in QCD vacuum. It has proven successful 
in extracting low-lying hadron masses and decay constants \cite{svz}. 
Several collaborations \cite{balitsky}-\cite{deviron} have studied
the masses of light hybrids with QCD sum rules.
The mass of heavy hybrid quarkonium has been studied by one group 
\cite{deviron-heavy}. But unfortunately the QCD sum rules for the 
exotic $(Q\bar Q g)$ states with $J^{PC}=0^{--}, 1^{-+}$ are not stable,
which prevents a reliable extraction of their masses \cite{deviron-heavy}.
Recently the masses and decay widths of hybrid mesons containing one heavy 
quark have been studied in the framework of HQET \cite{zhu}.

In this work we shall investigate the binding energy of hybrid quarkonium 
$(Q\bar Q g )$, the scaling behavior of their decay constants in the limit 
of $m_Q\to \infty$ and their $1/m_Q$ corrections. As we shall show later, 
all of our sum rules are stable after the heavy quark mass is separated out.

The interpolating current for the $J^{PC}=1^{-+}, 0^{++}$ heavy hybrid 
quarkonium reads
\begin{equation}
\label{curr1}
J_\mu (x) ={\bar Q}(x)g_s \gamma^\nu G^a_{\mu\nu}(x) {\lambda^a\over 2}
 Q (x)\;,
\end{equation}
and for the $J^{PC}=1^{+-}, 0^{--}$ ones
\begin{equation}
\label{curr2}
J^5_\mu (x) ={\bar q}(x)g_s \gamma^\nu\gamma_5 G^a_{\mu\nu}(x) {\lambda^a\over 2}
 h_v (x)\;.
\end{equation}

Denote the $J^{PC}=1^{-+}, 0^{++}, 1^{+-}, 0^{--}$ 
hybrid mesons by $H_i$ and the decay constants by $f_i$, $i=1, 2, 3, 4$.
For example, $f_1$ is defined as:
\begin{equation}\label{overlap-1}
\langle 0 | J_\mu (0) | H_1\rangle =f_1 \epsilon^1_\mu \; ,
\end{equation}
where $\epsilon^1_\mu$ is the $H_1$ polarization vector. 

We consider the correlators 
\begin{equation}\label{cor-1}
i\int d^4 x e^{ipx} \langle 0| T\{J_\mu (x), J^{\dag}_\nu (0)\} |0\rangle  
=-(g_{\mu\nu} -{p_\mu p_\nu\over p^2} )\Pi_1 (p^2)
+ {p_\mu p_\nu\over p^2}\Pi_2 (p^2) \;,
\end{equation}
\begin{equation}\label{cor-2}
i\int d^4 x e^{ikx} \langle 0| T\{J^5_\mu (x), J^{5 \dag}_\nu (0)\} |0\rangle  
=-(g_{\mu\nu} -{p_\mu p_\nu\over p^2})\Pi_3 (p^2)
+{p_\mu p_\nu\over p^2}\Pi_4 (p^2) \;.
\end{equation}
The imaginary parts of $\Pi_1, \Pi_2, \Pi_3, \Pi_4$ receive contributions 
from the $H_1, H_2, H_3, H_4$ hybrid intermediate states respectively.

The dispersion relation for $\Pi_i (p^2 )$ reads
\begin{equation}\label{dip-1}
\Pi ( p^2 )=\int {\rho (s)\over s- p^2 -i\epsilon }ds\;,
\end{equation}
where $\rho (s)$ is the spectral density.

At the phenomenological side
\begin{equation}\label{dip-2}
\Pi (p^2 )= {f^2
\over  M^2-p^2}+\mbox{excited states}+\mbox{continuum} \;.
\end{equation}

In order to derive the spectral density in full QCD we can either use the 
Cutkosky cutting rule or employ the dispersion relation and 
previous results of QCD sum rule for heavy quarkonium \cite{reinders}. 
With both methods we have obtained the same results independently.
For example, the spectral density in full QCD for $H_1$ reads
\begin{equation}\label{full}
\rho_1 (s)={\alpha_s \over 72\pi^3}\int_{4m_Q^2}^s 
\sqrt{1-{4m_Q^2\over x}}(1+{2m_Q^2\over x}) 
{(s-x)^3(s+x)\over s^2} dx 
+{<0|g_s^2G^2|0>\over 144\pi^2}
\sqrt{1-{4m_Q^2\over s}}(1+{2m_Q^2\over s}) \;,
\end{equation}
which agrees with the expressions in \cite{deviron-heavy}.

Note the gluon condensate and the perturbative term is of the same order
in the limit of $m_Q\to \infty$, which enables us
to explore the scaling behavior. Let $M=2m_Q +\Lambda$, where $\Lambda$ 
is the leading order binding energy. Expanding (\ref{dip-1}), (\ref{dip-2}), 
(\ref{full}) to the leading order of $1/m_Q$ and making Borel transformation 
to suppress the continuum contribution, we arrive at
\begin{equation}
\label{mass-1}
{1\over 4m_Q}f^2 e^{-{\Lambda\over T}} 
=\int_0^{E_c} m_Q^{3\over 2}
\rho_0 (\epsilon ) e^{-{\epsilon\over T}}d\epsilon\;,
\end{equation}
where $\rho_0 (\epsilon )=a_i {\alpha_s\over \pi^3}\epsilon^4\sqrt{\epsilon} 
+{b_i\over \pi^2}<0|g_s^2G^2|0> \sqrt{\epsilon}$, 
$\alpha_s ={4\pi\over (11-{2\over 3}n_f)
\ln ({E_c/2 \over \Lambda_{\small \mbox{QCD}}})^2}$
and $<0|g_s^2G^2|0> =0.48$ GeV$^4$. The coefficients $a_i$ etc 
are listed in TABLE I. $E_c$ is the continuum threshold. 
Starting from $E_c$ we have modeled the phenomenological spectral density 
with the parton-like one $m_Q^{3\over 2} \rho_0 (\epsilon )$. 
The triple gluon condensate is suppressed by a 
factor of $1/m^2_Q$ so it does not contribute up to the order ${\cal O}(1/m_Q)$. 
From (\ref{mass-1}) we can introduce a scaled decay constant 
$F=m_Q^{-{5\over 4}}f$, which remain constant in the limit $m_Q \to \infty$. 

We further expand the hybrid mass $M$ and decay constant $f$ 
to the order of $1/m_Q$, $M=2m_Q +\Lambda +{\Lambda_1\over m_Q}$, 
$f=m_Q^{5\over 4}(F+{F_1\over m_Q})$, $\Lambda_1$, $F_1$ is the 
${\cal O}(1/m_Q)$ correction to $\Lambda$ and $F$ respectively.
Similarly we have,
\begin{equation}
\label{mass-2}
{1\over 4}[2F_1F-{F^2(\Lambda_1 +{\Lambda^2\over 4})\over T}] 
e^{-{\Lambda\over T}} 
=\int_0^{E_c} \rho_1 (\epsilon ) e^{-{\epsilon\over T}}d\epsilon\;,
\end{equation}
where $\rho_1 (\epsilon )
=c_i {\alpha_s\over \pi^3}\epsilon^5\sqrt{\epsilon} 
+{d_i\over \pi^2}<0|g_s^2G^2|0> \epsilon\sqrt{\epsilon}$.

With (\ref{mass-1}) and (\ref{mass-2}) we get
\begin{equation}\label{mass-3}
\Lambda ={\int_0^{E_c} s \rho_0 (s) e^{-{s\over T}}ds \over
\int_0^{E_c} \rho_0 (s) e^{-{s\over T}}ds } \; ,
\end{equation}
\begin{equation}
\label{mass-4}
{F^2\over 4}( \Lambda_1 +{\Lambda^2\over 4})
=\int_0^{E_c} \rho_1 (\epsilon ) (\epsilon -\Lambda )
e^{-{\epsilon-\Lambda\over T}}d\epsilon\;,
\end{equation}
\begin{equation}
\label{FFF}
{F_1F\over 2}=\int_0^{E_c} \rho_1 (\epsilon ) 
(1+ {\epsilon -\Lambda \over T})
e^{-{\epsilon-\Lambda\over T}}d\epsilon\;.
\end{equation}

Requiring the absolute value of the gluon condensate be less than $30\%$ of 
the leading perturbative term with continuum subtracted we get the lower 
limit of the Borel parameter $T$ and the continuum threshold $E_c$. 
Typically in our analysis the gluon condensate contribution in the whole 
sum rule is less than $25\%$ for $1^{-+}$ and $10\%$ for the other three states
starting from $T >0.9$ GeV. We have kept four active flavors and let 
$\Lambda_{\mbox{QCD}}=220$ MeV. Varying $\Lambda_{\mbox{QCD}}$ from $220$ MeV 
to $300$ MeV the final result changes within $5\%$. 
We want to emphasize that all the sum rules for $\Lambda, \Lambda_1,
F, F_1$ are stable in the large interval $0.9 <T <3.0$ GeV. 

As a consistency check we have also fitted the left and right hand side 
of Eq. (\ref{mass-1}) directly with the most suitable parameters $ \Lambda, F, 
E_c$ directly in the working region of the Borel parameter. 
The results agree very well with those derived from the derivative method. 

We collect our final results in TABLE II. For $\Lambda$ 
there is an error about $(\pm 0.2)$ GeV. For the exotic state 
$H_1$ we present the variations of $\Lambda$ with $T$ and $E_c$
in FIG. 1. We also show the left and right hand side 
of (\ref{mass-1}) with the central values of $\Lambda, F, E_c$ in TABLE II
for $H_1$. Both sides agree within one percent in the 
region $0.5 < T < 3.0$ GeV as can be seen from FIG. 2. 

In the limit $m_Q\to \infty$ the $0^{--}$ and $0^{++}$ hybrid quarkonium 
is exactly degenerate while the ${\cal O}(1/m_Q)$ correction makes 
$0^{--}$ mass lower than that for $0^{++}$. The $1^{-+}$ and $1^{+-}$
hybrid is nearly degenerate and the ${\cal O}(1/m_Q)$ correction does 
not split their masses. The $1^{-+}$ hybrid lies $0.7$ GeV lower 
than the $0^{--}$ hybrid. The mass of the lowest exotic 
hybrid quarkonium $b\bar b g$ is $(10.75\pm 0.20)$ GeV if we use
$m_b=4.8$GeV.

In summary we have calculated the binding energy of the hybrid quarkonium
up to the order of ${\cal O}(1/m_Q)$. We have found that the decay constants
of the hybrid mesons scale like $m_Q^{5\over 4}$ as $m_Q\to \infty$.
Moreover, the scale of the binding energy is solely set by the gluon condensate.
In the sum rules for the light hybrids the four quark condensate 
and three gluon condensate are not precisely known, which renders 
the extraction of the light hybrid mass rather difficult 
\cite{balitsky,latorre}. 
In our calculation of hybrid quarkonium masses we have not considered 
the error due to the bottom quark mass. Another possible source of
error is the lack of $E_c$ stability. 

\vspace{0.8cm} {\it Acknowledgments:\/}This work was supported by
the Natural Science Foundation of China. The author is grateful to 
Prof. J. Govaerts for helpful communications.

\newpage 
{\bf Figure Captions}
\vspace{2ex}
\begin{center}
\begin{minipage}{130mm}
{\sf FIG. 1} 
\small{The variations of $\Lambda$ with $T$ and $E_c$ for $H_1$. 
$T$ is in unit of GeV.}
\end{minipage}
\end{center}
\begin{center}
\begin{minipage}{130mm}
{\sf FIG. 2} \small{The variation of the right and left hand side 
of Eq. (\ref{mass-1}) with $T$ is plotted as solid and dotted curves 
respectively for $H_1$ with the values in TABLE II. }
\end{minipage}
\end{center}
\vspace{2.0cm}

TABLE I. The coefficients $a_i$ etc in $\rho_0 (\epsilon )$ and $\rho_1 (\epsilon )$.
\vskip 0.5cm
\begin{tabular}{|c|c|c|c|c|}
\hline
 & $1^{-+}$ & $0^{++}$ &$1^{+-}$  & $0^{--}$   \\
\hline
a &${256\over 945}$& ${384\over 945}$&${128\over 405}$&${256\over 945}$  \\ 
b &${1\over 24}$&$-{1\over 16}$&${1\over 144}$&$-{1\over 24}$  \\ 
c &$-{128\over 945}$& $-{3648\over 10395}$&$-{1216\over 10395}$&$-{2176\over 10395}$  \\ 
d &${1\over 144}$&$-{1\over 96}$&${7\over 288}$&$-{1\over 48}$  \\ 
\hline
\end{tabular}
\vskip 2.0 true cm

TABLE II. The values of $\Lambda, \Lambda_1, F, F_1, E_c$ for the hybrid quarkonium.
$\Lambda, E_c$ is unit of GeV, $\Lambda_1$ is unit of GeV$^2$,
$F$ is in unit of GeV$^{11\over 4}$ 
and $F_1$ is in unit of GeV$^{15\over 4}$.
\vskip 0.5cm
\begin{tabular}{|c|c|c|c|c|}
\hline
 & $1^{-+}$ & $0^{++}$ &$1^{+-}$  & $0^{--}$   \\
\hline
$\Lambda$ &$1.22$& $1.97$&$1.21$&$1.97$  \\ 
$\Lambda_1$ &$-0.44$& $-0.44$&$-0.42$&$-0.92$  \\ 
$F$ &$0.244$& $0.89$&$0.208$&$0.515$  \\ 
$F_1$ &$-0.064$& $-0.105$&$-0.036$&$-0.42$  \\ 
$E_c$ &$1.6$& $2.3$&$1.5$&$2.3$  \\ 
\hline
\end{tabular}
\vskip 1.0 true cm

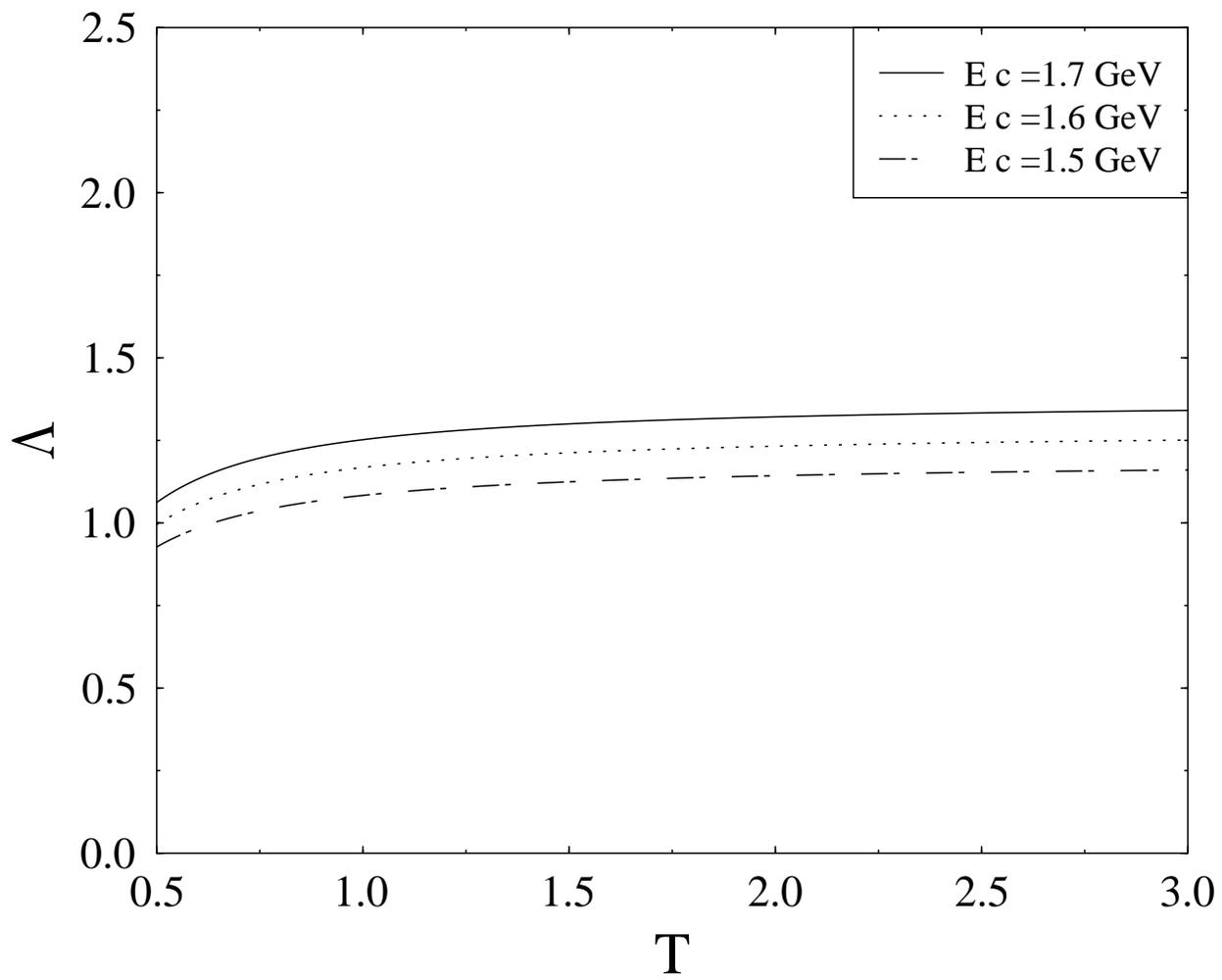
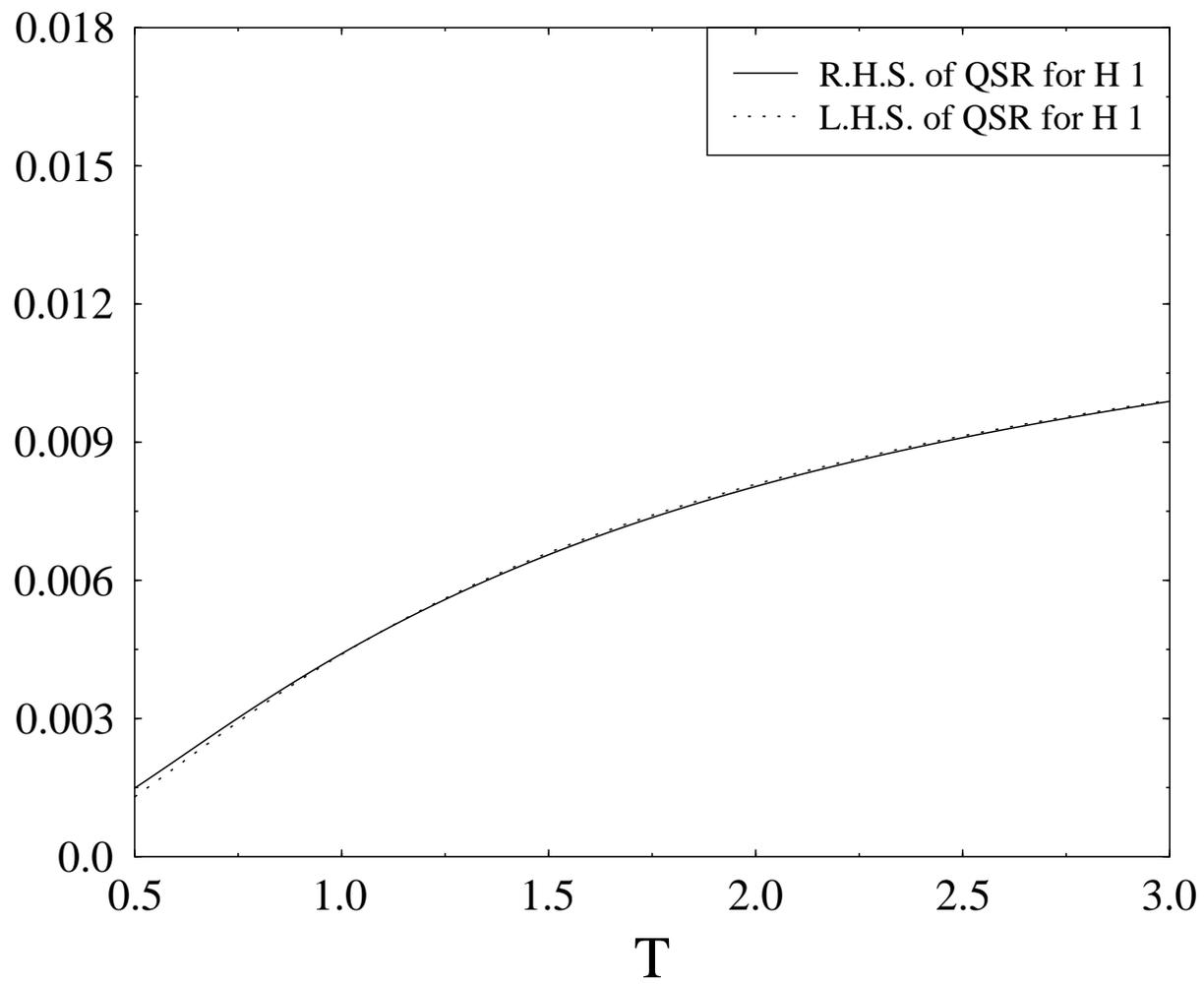
\end{document}